\documentclass[useAMS,usenatbib]{mn2e}
\usepackage{times}
\input{psfig}
\def\<#1>{\langle\hbox{#1}\rangle}
\def\etal{{et~al.}}
\def\kms{km~s$^{-1}$}

\def\eg{{e.g.}}

\def\simgt{\hbox{\rlap{\raise 0.425ex\hbox{$>$}}\lower 0.65ex\hbox{$\sim$}}}
\def\simlt{\hbox{\rlap{\raise 0.425ex\hbox{$<$}}\lower 0.65ex\hbox{$\sim$}}}

\def\deriv(#1/#2){\mathchoice{d#1\over d#2}
    {d#1/d#2} {d#1/d#2} {d#1/d#2}}
\def\pderiv(#1/#2){\mathchoice{\partial#1\over\partial#2}
    {\partial#1/\partial#2} {\partial#1/\partial#2} {\partial#1/\partial#2}}
\def\pderivp(#1/#2){\left(\pderiv(#1/#2)\right)}

\def\tderiv(#1/#2){\mathchoice{d#1\over d#2}{d#1/d#2}
                   {d#1/d#2}{d#1/d#2}}
\def\tderivp(#1/#2){\left(\tderiv(#1/#2)\right)}
\def\<#1>{\langle\hbox{#1}\rangle}
\def\etal{{et~al.}}
\def\kms{km~s$^{-1}$}

\def\eg{{e.g.}}

\def\simgt{\hbox{\rlap{\raise 0.425ex\hbox{$>$}}\lower 0.65ex\hbox{$\sim$}}}
\def\simlt{\hbox{\rlap{\raise 0.425ex\hbox{$<$}}\lower 0.65ex\hbox{$\sim$}}}
 
\def\deriv(#1/#2){\mathchoice{d#1\over d#2}
    {d#1/d#2} {d#1/d#2} {d#1/d#2}}
\def\pderiv(#1/#2){\mathchoice{\partial#1\over\partial#2}
    {\partial#1/\partial#2} {\partial#1/\partial#2} {\partial#1/\partial#2}}
\def\pderivp(#1/#2){\left(\pderiv(#1/#2)\right)}

\def\tderiv(#1/#2){\mathchoice{d#1\over d#2}{d#1/d#2}
                   {d#1/d#2}{d#1/d#2}}
\def\tderivp(#1/#2){\left(\tderiv(#1/#2)\right)}
\begin{document}
\title[Star formation in galaxies falling into clusters]
   {Star formation in galaxies falling into clusters
          along supercluster-scale filaments}
\author[Porter, Raychaudhury, Pimbblet \& Drinkwater]{Scott C. Porter$^{1}$,
Somak Raychaudhury$^{1}$\thanks{Corresponding author:
              somak@star.sr.bham.ac.uk},
Kevin A. Pimbblet$^{2}$ \& Michael~J.~Drinkwater$^{2}$\\
\vspace*{6pt} \\
$^{1}$School of Physics and Astronomy, University of Birmingham,
Birmingham B15~2TT, UK\\
$^{2}$Department of Physics, University of Queensland, Brisbane,
             QLD 4072,  Australia}

\pagerange{\pageref{firstpage}--\pageref{lastpage}} \pubyear{2007}

\maketitle

\label{firstpage}

\begin{abstract}
With the help of a statistical parameter derived from optical spectra,
we show that the current star formation rate of a galaxy, falling into
a cluster along a supercluster filament, is likely to undergo a sudden
enhancement before the galaxy reaches the virial radius of the
cluster.  From a sample of 52 supercluster-scale filaments of galaxies
joining a pair of rich clusters of galaxies within the two-degree
Field Redshift Survey region, we find a significant enhancement of
star formation, within a narrow range between
$\sim$2--3$\,h_{70}^{-1}$~Mpc of the centre of the cluster into which
the galaxy is falling.  This burst of star formation is almost
exclusively seen in the fainter dwarf galaxies ($M_B \!\ge\! 
-20$). The relative position of the peak does not depend on whether
the galaxy is a member of a group or not, but non-group galaxies have
on average a higher rate of star formation immediately before falling
into a cluster. From the various trends, we conclude that the
predominant process responsible for this rapid burst is the close
interaction with other galaxies falling into the cluster along the
same filament, if the interaction occurs before the gas reservoir of
the galaxy gets stripped off due to the interaction with the
intracluster medium.
\end{abstract}

\begin{keywords}
Galaxies: clusters: general;  Galaxies: evolution; large-scale
structure of the Universe -- cosmology: observations; Galaxies: starburst.
\end{keywords}

%

\section{Introduction}

In the past few years, several examples of spectacular starburst
galaxies, often with significantly disturbed morphology, have been
discovered on the outskirts of rich clusters of galaxies. 
Most of these have been found
in deep images or spectral line studies of galaxies in rich clusters
at redshifts of $z\!=$0.2--0.4
\citep{moran2005,sm06,braglia07,marcillac07,einasto07,fadda2008}.  
Some of these have been
found in studies of nearby ($z\!<\!0.1$) clusters as well
\citep{sun02,gav03,haines06a,reverte07}
though in the absence of systematic studies of large samples, it is
not clear how common they are, and whether their presence is related
to special properties of the clusters they are observed in, or to
their large-scale environment, or indeed to the properties of the
galaxies themselves.

The evolution of a galaxy is profoundly affected by its
environment. Observationally, this has been established by studying
the the dependence of the relative morphological content of galaxies
in a given volume of space, or of the colour or 
star formation properties of
galaxies, on the local projected galaxy density. An obvious place to
look for such an effect is in galaxy
clusters, which provide a wide range of galaxy densities as well
as individual galaxy properties.

In the cores of rich clusters, the current star formation rate (SFR)
in galaxies is found to be 
much lower than in the field, as measured from
colour, morphology or spectral lines,
progressively increasing as the environment becomes less dense
\citep[\eg][]{ms77,dressler1980,
balogh02, lewis2002,gomez2003,deprop2003,kauf2004,
baldry06,pog06,pimbblet2006,verdugo07}.
Whether the environment should be
quantified by only the local projected density or not
has also been the subject of
scrutiny \citep[\eg][]{kauf2004,blaber07}. Given that galaxies and
groups are embedded in a network of clusters and
filaments \citep{einasto94,virgo01}, an interesting aspect of such
studies is the investigation of how important the large-scale
properties of the environment of a galaxy is in its morphological
evolution, as well its history of star formation, merger and nuclear
activity.  Even though mergers and close interactions with other galaxies
are most influenced by the local group environment of most
galaxies \citep[\eg][]{hopkins07},
the potential effect of the larger-scale environment is
undeniable, given that it would fundamentally affect the history of
the parent group of the galaxy.

Recently, \citet{porter2007}  investigated the environmental dependence of
star formation within three filaments (of length
between 20--30$\,h^{-1}_{70}$ Mpc), joining rich clusters of the
Pisces-Cetus supercluster.
We observed that as one moves away
from the cluster cores there is an increased activity of star
formation, with a peak in the 3-4~$h_{70}^{-1}$~Mpc range along the
filaments over and above the gradual increase in star formation rate
(SFR) away from the cluster cores. While this is a striking result,
these results come only from approximately 1,000 galaxies in three
filaments, and in one supercluster. To confirm that the observed
effect is not confined to this one supercluster, a larger sample of
filaments is therefore required. 
The Pisces-Cetus supercluster is the only ``supercluster''
found in the 2dFGRS survey from a study of all Abell clusters
with a redshift of 0.1 from a minimal 
spanning tree analysis \citet{somak08} catalogue. To enlarge this 
sample of galaxies, we seek a larger survey of filament-like structures.

In this paper, we have compiled a subset
of ``clean filaments'' (defined below)
from the filament catalogue of
\citet{pimbblet2004} which listed 805 filaments identified from the
galaxies of the 2dFGRS.  We describe this sample in
\S\ref{sec:sample}, and in 
investigate the star formation properties of galaxies belonging to
these filaments as a function of distance from the nearest cluster,
We discuss our results in
\S\ref{sec:dis} and our conclusions can be found in
\S\ref{sec:con}. 
Throughout this paper we use $H_0\!=\!70$~km~s$^{-1}$ Mpc$^{-1}$,
and a $\Omega_M=1$ CDM cosmology. Using a concordance
cosmology with a dark energy component affect only our
distance measures, but not appreciably over the volume
covered by our analysis.

\begin{figure*}
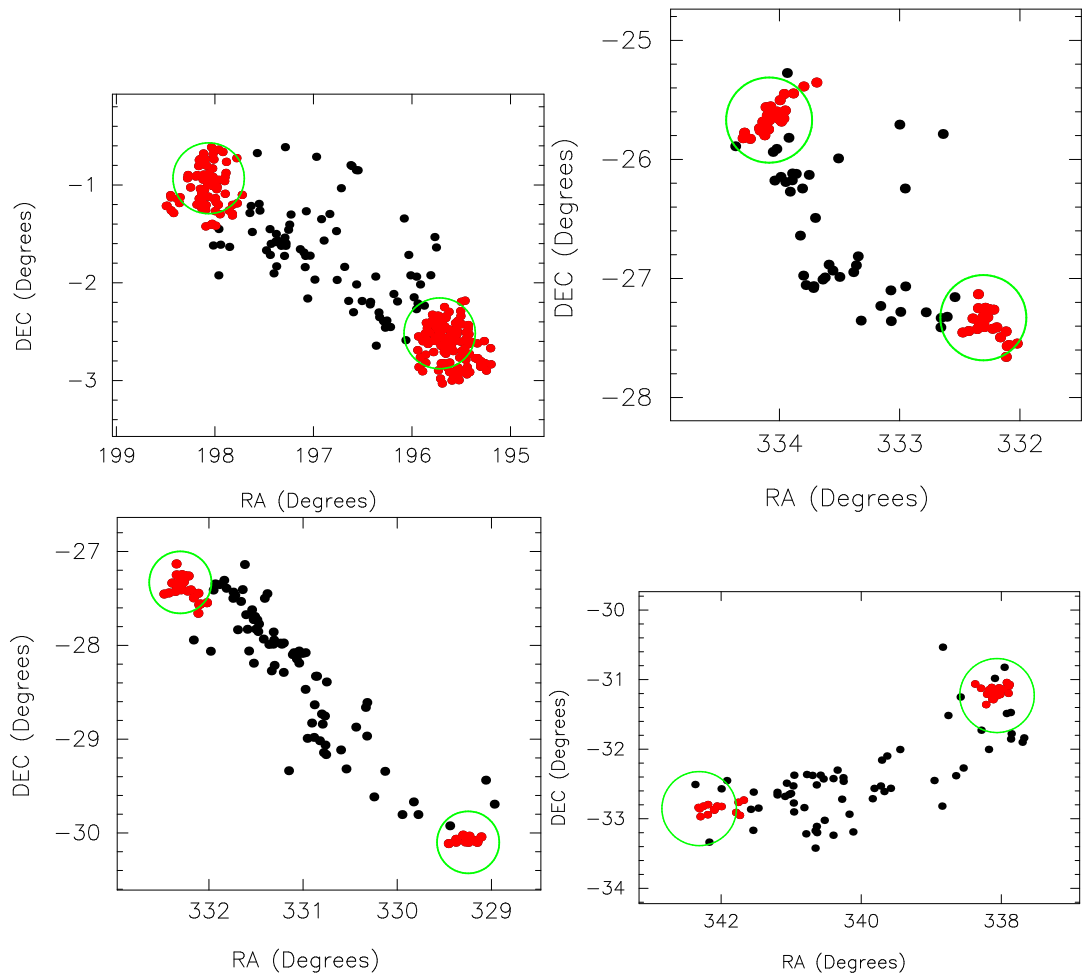

\centerline{
\psfig{figure=2df-fig1a.ps,width=0.4\hsize,angle=-90}
\psfig{figure=2df-fig1b.ps,width=0.4\hsize,angle=-90}}
\centerline{
\psfig{figure=2df-fig1c.ps,width=0.4\hsize,angle=-90}
\psfig{figure=2df-fig1d.ps,width=0.4\hsize,angle=-90}
}
\medskip
 \caption{A few of the ``clean filaments''
used in this paper. The dots represent the galaxies
from the 2dFGRS (with spectroscopic redshifts)
included in these filaments joining the clusters
(a) Abell 1692 and Abell 1663, (b) Abell 3837 and EDCC 0119,
(c) Abell 2660 and APMCC 917, and 
(d) EDCC 365 and Abell S1155. 
 Cluster members are shown in red, and the Abell
radius ($2.1\,h_{70}^{-1}$Mpc) is shown as a green circle.
\label{fill12}}
\end{figure*}


\section{The sample of filaments}
\label{sec:sample}

Our parent filament catalogue 
\citep{pimbblet2004}
is sourced from the 2 degree field galaxy redshift survey (2dFGRS;
Colless et al., 2001) which in turn is based upon the APM survey
\citep[e.g.,][]{maddox90} as its photometric input catalogue.  Here we
summarise the most pertinent points, directing the interested reader
to the filament catalogue paper for more detail.

2dFGRS galaxies are selected from the APM for observation to have an 
extinction corrected magnitude of $b_J\!<\!19.45$ 
Using the final data release of 2dFGRS and combining cluster 
information 
\citep{deprop2002,pimbblet2004} examined by 
the the region between $>$800 close cluster pairs (where clusters are
within 10 degrees of each other and their difference in mean recession 
velocity is $<$1000 \kms.

\begin{table}
\centering
\begin{minipage}{0.5\textwidth}
\centering
\caption{The 52 inter-cluster filaments for the 2dFGRS
survey used in this study. \label{fillst}} 
\begin{tabular}{lcccc} 
\hline 
{Clusters}&{Mean}&{Filament}&EETMA\\
&redshift&length &Id\\
&&($h_{70}^{-1}$ Mpc)&\\
\hline
Abell  2829 Abell  0118 & 0.1133 & 20.75 & 9\rlap{$^a$} & \\
Abell  2780 APMCC  0039 & 0.1031 & 33.62 & 9 &  \\
EDCC   0465 Abell  2780 & 0.1046 & 40.13 & 9 &  \\
Abell  2814 Abell  2829 & 0.1097 & 22.08 & 9 &  \\
Abell  2829 EDCC   0511 & 0.1112 & 23.01 & 9 &  \\
EDCC   0492 Abell  2804 & 0.1169 & 44.83 & 9 &  \\
Abell  3094 Abell S0333 & 0.0675 & 13.07 & 49 &  \\
Abell  1419 Abell  1364 & 0.1072 & 27.51 & 100,265 &  \\
Abell  1411 Abell  1407 & 0.1334 & 21.41 &  &  \\
Abell  1692 Abell  1663 & 0.0834 & 17.41 &  126&  \\
Abell  1620 Abell  1663 & 0.0839 & 21.58 &  126&  \\
Abell  2553 EDCC   0275 & 0.1458 & 13.05 &  &  \\
Abell  3980 EDCC   0268 & 0.1894 & 27.04 &  &  \\
APMCC  0869 APMCC  0840 & 0.1125 & 47.62 &  &  \\
Abell  2601 Abell  4009 & 0.1091 & 42.68 &  299&  \\
EDCC   0365 Abell S1155 & 0.0547 & 38.71 &  &  \\
Abell  2741 APMCC  0039 & 0.1050 & 12.92 &  &  \\
Abell  2741 APMCC  0051 & 0.1060 & 21.07 &  &  \\
EDCC   0445 APMCC  0094 & 0.0617 & 17.94 &  &  \\
APMCC  0094 EDCC   0457 & 0.0613 & 20.86 &  &  \\
EDCC   0465 Abell  2814 & 0.1084 & 12.44 &  &  \\
Abell  2734 EDCC   0445 & 0.0623 & 17.94 &  10\rlap{$^b$}&  \\
EDCC   0457 EDCC   0445 & 0.0620 & 10.24 &  &  \\
EDCC   0517 EDCC   0511 & 0.1107 & 19.03 &  &  \\
Abell  2878 EDCC   0511 & 0.1090 & 24.14 &  &  \\
Abell  2915 EDCC   0581 & 0.0862 & 23.51 &  28&  \\
APMCC  0167 Abell S0160 & 0.0688 &  7.98 &  &  \\
Abell  2943 APMCC  0222 & 0.1498 & 21.36 &  &  \\
Abell  2961 APMCC  0245 & 0.1242 & 25.60 &  232&  \\
Abell  2967 Abell  2972 & 0.1122 & 12.03 &  232&  \\
Abell  2967 Abell  2981 & 0.1103 & 16.12 &  232&  \\
Abell  2981 Abell  2999 & 0.1083 & 15.84 &  232&  \\
EDCC   0119 Abell  3837 & 0.0885 & 28.84 &  190&  \\
EDCC   0057 Abell  3837 & 0.0922 & 25.07 &  190&  \\
EDCC   0057 APMCC  0721 & 0.0963 & 30.55 &  &  \\
Abell  3878 Abell  3892 & 0.1179 & 22.31 &  &  \\
EDCC   0230 APMCC  0827 & 0.1103 & 15.80 &  &  \\
APMCC  0827 Abell  3892 & 0.1142 & 31.64 &  &  \\
EDCC   0128 Abell  3880 & 0.0586 & 15.12 &  199&  \\
Abell  3959 APMCC  0853 & 0.0876 & 17.09 &  &  \\
Abell S1075 Abell  3978 & 0.0854 & 26.87 &  &  \\
EDCC   0457 Abell  2794 & 0.0613 & 20.86 &  &  \\
EDCC   0202 EDCC   0187 & 0.0768 &  8.85 &  &  \\
EDCC   0187 APMCC  0810 & 0.0769 &  7.24 &  &  \\
EDCC   0317 Abell  4011 & 0.1371 &  8.76 &  &  \\
EDCC   0321 Abell  4012 & 0.0533 &  9.49 &  &  \\
Abell  3854 Abell  3844 & 0.1505 & 29.89 &  &  \\
Abell S1064 EDCC   0153 & 0.0571 & 17.95 &  &  \\
EDCC   0239 APMCC  0853 & 0.0871 & 17.62 &  &  \\
EDCC   0248 Abell  3959 & 0.0876 & 17.09 &  &  \\
EDCC   0217 EDCC   0202 & 0.0784 & 19.94 &  &  \\
Abell  2493 EDCC   0215 & 0.0773 & 21.88 &  &  \\
\hline
\end{tabular}
\end{minipage}
\noindent{ $^a$Sculptor Supercluster; $^b$Pisces-Cetus Supercluster (from 
\citet{einasto01}.} 
\end{table}

Further criteria were applied to narrow down the initial
\citet{pimbblet2004} sample of 805 potential filaments.  Only
filaments having a length of between 10 and 40~$h_{70}^{-1}$~Mpc were
selected.  Further more, potential filaments that had a classification
of ``near coincidental clusters'' or ``Nil'' were also removed.  This
resulted in a sample of 432 filaments to study visually.  Each of the
432 filaments had two clusters from which \citet{pimbblet2004} had
identified a filament.  In addition to these clusters other clusters
fall within the same region of space.  These extra clusters were
deemed to be part of the filament if they had a redshift within 0.01
of the mean redshift of the two initial clusters and were coincident
with the filament galaxies in R.A. and Dec. space.  In addition, to
confirm the presence of a cluster of galaxies at the cluster
positions, each of the clusters had to have greater than ten members
to be part of the filament.  Cluster membership was determined by
finding all 2PIGG \citep{eke2004} group centres within
1~$h_{70}^{-1}$~Mpc of the cluster centres. The members of these
groups were then taken as the cluster members. Most clusters would
correspond with a single 2PIGG ``group'' but for some, the 2PIGG
catalogue had separated the cluster members into multiple ``groups''.

From the sample of 432 filaments we created a ``clean sample'' of
filaments. The filaments had to be between two clear clusters of
galaxies and have to be relatively clean from contamination from other
filaments and clusters.  Therefore, by a visual inspection of the 432
filaments, 52 pairs of clusters on these filaments that had a clear
run of galaxies, with no other intruding clusters within
6$\,h_{70}^{-1}$~Mpc, were found.  To avoid contamination from the
complex weave of other filaments, and to enable comparisons with our
previous work, the galaxies joining the two filament clusters were
selected using a method which extracts galaxies within a prolate
spheroid.

We model each filament as a prolate spheroid, with the centres of the
two clusters involved being at each end, the distance between them
being the major axis of the spheroid, and $6\,h_{70}^{-1}$ Mpc being
the semi-minor axis.  All galaxies inside this spheroid are assumed to
be members of the filament, along the galaxies found to belong to the
2PIGG ``group'' corresponding to these Abell clusters in
\citep{eke2004}, to avoid leaving out the genuine cluster members that
would not be included in the spheroid.

The distance between each pair of
clusters, or that between a galaxy and a cluster,
was calculated as the comoving proper distance between them. 
We used the measured redshift of each
galaxy as a measure of its distance from us, except if it is a member
of a rich cluster, where the mean redshift of the cluster was used.

This resulted in 6,222 galaxies within the 52 filaments.  Examples of
the spatial structure of the these filaments can be seen in
Figs.\ref{fill12}, and a complete list of the filaments, and the
clusters they connect, can be found in Table~\ref{fillst}. Apart from
Abell clusters, the cluster catalogue used to find filaments also use
supplementary Abell (Abell~S), Edinburgh-Durham (EDCC) and APM
clusters (APMCC) within the same redshift range
\citep{aco89,edcc92,apmcc}. 
All references to 2dFGRS galaxies are taken from the 2PIGG subsample
of the 2dFGRS \citep{eke2004} comprising 191,328 galaxies.  This
subsample had all the galaxies within fields with less than 70$\%$
completeness, and in all sectors (overlapping field areas) with
completeness less than 50$\%$, removed This enables comparisons with
2PIGG statistics to be made without bias.

In the Table, we also list the
corresponding supercluster in the \citet[EETMA]{einasto01} catalogue,
resulting from their analysis of the supercluster-void network from
Abell rich clusters.  Not all of our filaments correspond to their
superclusters because of the different algorithm and cluster samples
used in identifying filaments. It is interesting to note that the
Sculptor supercluster, one of the most richest superclusters in the
\cite{einasto01} study, features prominently in this Table.


\section{Star formation in galaxies along the filaments}
\label{sec:sfr1}

The star formation properties of the 6,222 galaxies in the ``clean sample''
of filaments are investigated in this section. 
We look at their rate of star formation (represented by the $\eta$ parameter)
and the ratio of passive to star-forming galaxies, as a function of distance 
along the filaments.

To quantify the star formation properties of these galaxies, we 
use a simple parameter
that has
been derived for most 2dFGRS galaxies.
\citet{madgwick2002} used Principal Component Analysis
\citep[also see][]{folkes99}
of de-redshifted 2dF galaxy spectra to define a parameter $\eta$, a
linear combination of the first two principal components, which
correlates well with the equivalent 
width of the $H\alpha$ (EW($H\alpha$)) emission
line, which in turn is a measure of SFR
\citep[e.g.,][]{ken1983,gall1984,moustakas2006}. 
With some scatter,
$\eta\approx -2.0$ correspond to no $H\alpha$ emission at all,
increasing to 
$\eta\approx 7$
for EW($H\alpha$) of 50 \AA\
\citep{madgwick2003}. Thus, the $\eta$ parameter
can be used as a proxy for star formation rate in galaxies
\citep[see][]{porter2005,porter2007}.

\begin{figure}
\centerline{
\psfig{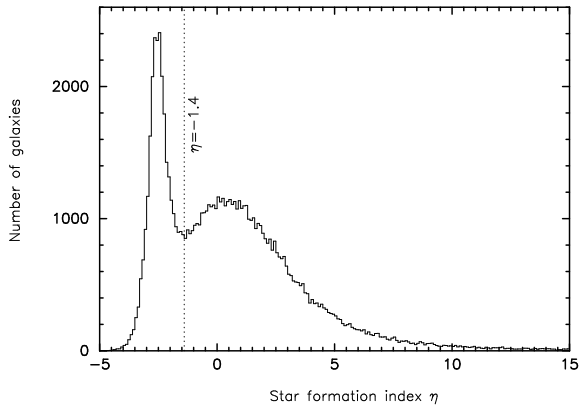}}
\medskip
\caption[]{\footnotesize
A histogram of the $\eta$ parameter for all 2dFGRS galaxies within $z<0.1$.
The vertical dashed line is our adopted division between
star-forming (larger $\eta$) and passive galaxies.
\label{2dfhist1}}

\end{figure}

Fig.~\ref{2dfhist1} shows 
the distribution of the $\eta$ parameter for all 2dFGRS galaxies
within $z\!\le \!0.1$,
revealing
two distinct peaks in the distribution, at
$\eta\!\sim \!-2.5$ and at $\eta\!\sim\! 0.05$. This reflects the
well-known bimodality of the local galaxy population, of red, passive
galaxies, and bluer, star forming galaxies
\citep[\eg][]{kauf2004,balogh04}.  The dip or divide between the two
peaks is approximately at $\eta\!\sim\! -1.4$. The first peak,
containing $\sim$30$\%$ of the galaxies within $z\!<\! 0.1$,
consists mostly of early-type galaxies
\citep{madgwick2003}.
The SFR of a galaxy 
normalised
to the Schechter luminosity
$L_\ast$, is shown to be $\mu_\ast\!=\!0.087$~EW(${H_\alpha}$), where the
EW(${H\alpha}$) is the stellar absorption corrected equivalent width
of the $H\alpha$ line
\citep{lewis2002}. Therefore, an $\eta\sim -1.4$
would correspond to $\mu_\ast\sim 0.4$.

\subsection{Star formation in galaxies as a function of distance
from the nearest cluster}

In this study of the star formation properties of galaxies, we look at the
variation of 
(a) the mean value of 
the $\eta$ parameter, which correlates with the current SFR of the galaxy,
and (b) the fraction of galaxies with $\eta < -1.4$, 
representing the fraction of passive galaxies,
as a function of position along the filaments
and distance away from the nearest cluster.

\begin{figure}
\centerline{
\psfig{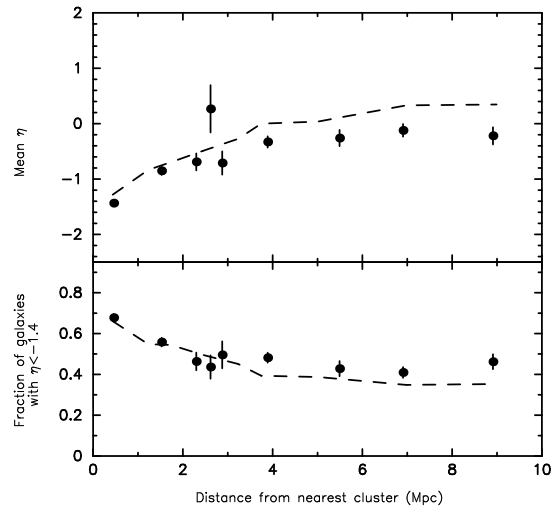}}
\medskip
 \caption[Mean $\eta$ as function of distance from the nearest
 cluster.]  {\footnotesize{ (a, Top) 
 The mean value of $\eta$ of galaxies belonging to
the ``clean'' sample of 52 filaments,
  as a function of distance from
 the nearest cluster, is plotted here.  
 The dashed line shows the mean $\eta$
 as function of distance from the nearest cluster (defined as a 2PIGG
 group with $\ge 30$ members) for {\em all} 
 2dFGRS galaxies.  (b, Bottom)
 For the same galaxies as in the top panel the fraction of these
 galaxies with an $\eta<-1.4$ is shown as a function of distance from
 the nearest cluster.  The dashed line showing the same fraction as a
 function of distance from the nearest 2PIGG group with $\ge 30$
 members for the whole 2dFGRS.}}
\label{sfr1}
\end{figure}

For each of the
6,222 galaxies  of the ``clean sample'' filaments, 
the distance of each from the nearest of the two
clusters at either end of their parent filament
is determined. The distances were then binned in varying size
bins, aiming for approximately equal numbers of 
galaxies in each bin, and
within each bin the mean $\eta$ was calculated. The data for galaxies
from all the filaments were combined and the means plotted against the
mean distance in each bin and can be seen in see Fig.~\ref{sfr1}.  For
comparison between the trend with distance for the filament galaxies
with galaxies elsewhere, we compute ``field'' values from the whole
2dFGRS, where distances are calculated from the nearest 2PIGG group of
$\ge 30$ members (the equivalent of a rich cluster), shown as the
dashed line.

In order to statistically compare the two sets of points in
Fig.~\ref{sfr1}, we use Welch's $t$-test \citep{welch47}, also
known as the $F$-test \citep{nrasbook}, which is an adaptation
of the Student's $t$ test, for the comparison
of two samples $x_A$ and $x_B$ with unequal variance
$\sigma_A^2\ne \sigma_B^2$.
This statistic 
tests  the null hypothesis that the means of the two samples
are equal, assuming a normally distributed parent
population, and returns a statistic 
$$t=\frac{\langle x_A\rangle -\langle x_B\rangle}{\LARGE[
\frac{\sigma_A^2}{N_A} +\frac{\sigma_B^2}{N_B}\LARGE]^{1/2}},$$
where $N_A$ and $N_B$ are the number of data points in the two
samples. The test also
 a value for the significance of the difference of the two
means, which is a number between 0 and 1.

In each of the 9 distance bins in Fig.~\ref{sfr1}, 
where the mean value of $\eta$ is presented for two  
samples (the galaxies belonging to our filaments, and
all galaxies in the 2dFGRS catalogue), we calculate the
probability,
that the means of the two samples are the same,
in each distance bin. The results are
shown in 
Table~\ref{tab:welch}, where it is obvious that the
two samples are similar in the first two bins, i.e.
closer than $2 \,h_{70}^{-1}$ Mpc from the centre of each cluster.
The galaxies on the inter-cluster filaments deviate appreciably
from the general galaxy sample from the fourth bin onwards.

It can be seen that while the field galaxies show a gradual decrease
in SFR as one moves from the periphery of the filament towards the
cluster, steepening from approximately $2 \,h_{70}^{-1}$ Mpc inward,
the filament galaxies show a sudden enhancement in SFR at about $2
\,h_{70}^{-1}$ Mpc on top of this gradual decline. 
This is very similar
to the effect seen in the three filaments of the Pisces-Cetus
supercluster, from a much smaller sample of galaxies \citep{porter2007}.

\begin{table*}
\caption{Welch's $t$-test results \label{tab:welch}} 
\begin{tabular}{lccccccccc} 
\hline 
Figure&
\multicolumn{9}{|c|}{Probability 
that the means of the two samples are the same in distance bin}\\
Number& 1&2&3&4&5&6&7&8&9\\
\hline 
Fig.~\ref{sfr1}&0.02&0.87&0.97&0.08&0.03&0.04&0.06&3.9E-3&6.5E-3\\
Fig.~\ref{sfr1dg}&1.1E-6&6.2E-10&0.24&0.001&0.04&3.9E-13&2.5E-10&1.5E-9&0.02\\
\hline
\end{tabular}
\end{table*}

\subsection{Star formation as a function of galaxy luminosity}

It has been shown that the SFR in giant and dwarf galaxies 
show very different properties \citep[e.g.,][]{haines06b}
and indeed, enhancements in SFR have been seen in dwarf galaxies
at approximately the virial radius \citep[e.g.,][]{moran2005}.
In \citet{porter2007}, we found that the peak in star
formation 
along the Pisces-Cetus filaments
is entirely due to the dwarf galaxies ($-20\!<\! M_B \!\le\! -17.5$). 

\begin{figure}
\centerline{
\psfig{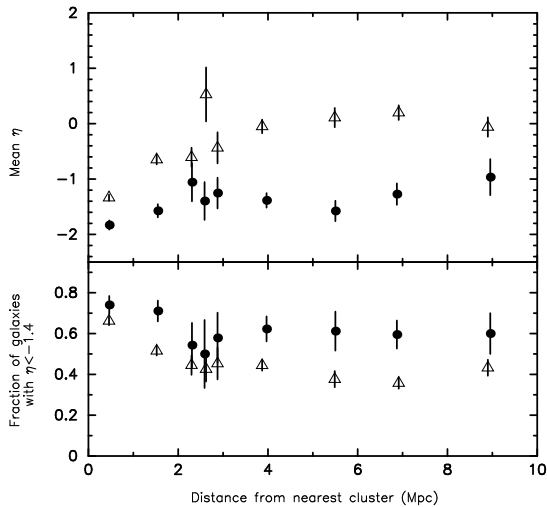}}
\medskip
 \caption[Mean $\eta$ of giant or dwarf galaxies as a function of
 distance from the nearest cluster.]  {\footnotesize{ (a, Top) For the
 ``clean filament'' sample (N=52), the mean value
 $\eta$ for giant galaxies ($M_B\!<\!-20$, closed circles) 
 and dwarf galaxies ($M_B\!>\!-20$, open triangles), is plotted 
 as a function of
 distance from the nearest cluster.
  (b, Bottom) The fraction of galaxies (same symbols) with
  $\eta<-1.4$, as a function of distance from nearest cluster.}
}
\label{sfr1dg}
\end{figure}

Consequently, our sample of 6,222 galaxies was
further segregated by environment into 1,392 
giant galaxies ($M_B \!\le\! -20$) and 4,830 dwarf
galaxies ($-20\!<\! M_B \!\le\!  -17.5$) respectively.
As before we plot the mean value of $\eta$, and the fraction of
passive galaxies, as a function of their distance from the nearest
cluster, which can be seen in Fig.~\ref{sfr1dg}. 
However, the two
separate samples are now plotted as open triangles for dwarf galaxies
and filled circles for giant galaxies.

In each of the distance bins in Fig.~\ref{sfr1dg}, 
where the mean value of $\eta$ is presented for two  
samples,
the ``dwarf'' ($M_B\!>\!-20$) and giant ($M_B\!<\!-20$)
galaxies, we calculate the
probability that the means of the two sample are
the same, according to the Welch test, 
in each distance bin. The results are
shown in 
Table~\ref{tab:welch}, where it is obvious 
the two sample are significantly different throughout
the entire distance range.

It can be seen that the dwarf galaxies have a higher mean SFR than
the giants at all values of 
distance from the nearest cluster. The striking revelation
in this plot is that 
the abrupt enhancement in SFR,
around a cluster-centric distance of about 2~Mpc,
seen in Fig.~\ref{sfr1}, is
almost entirely due to the dwarf galaxies.

\begin{figure}
\centerline{\psfig{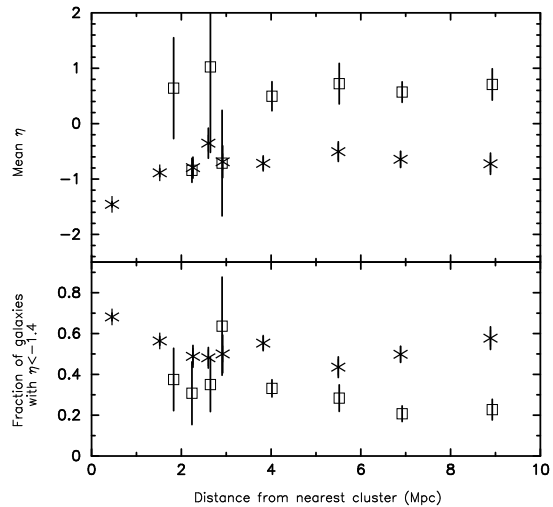}} 
\caption[]{\footnotesize {(a, Top) For the ``clean filament'' sample
  (N=52), the mean value of 
  $\eta$ as a function of distance from the nearest
  cluster.  Galaxies that are members of groups (2PIGG groups
  with $\ge 4$ members) and part of the
  filament are shown as crosses, while non-group galaxies
  are shown as squares. 
  (b, Bottom) The fraction of galaxies (same symbols) with
  $\eta<-1.4$, as a function of distance from nearest cluster.
}\label{sfrg}}
\end{figure}

\subsection{Star formation in galaxies belonging to Groups}
\label{sec:sfrg}

It has been suggested that much of the evolution of galaxies occurs in
groups during their life on the filaments, before the groups are
assimilated in clusters.  This can be seen when the suppression of
star formation with increasing local density of galaxies has been
observed to occur at very low values of the local projected density
($\sim$1~$h_{70}^{-1}$~Mpc$^{-2}$ \citep[\eg][]{lewis2002,gomez2003}.
If this is the case the trend seen in the previous section would be
different for filament galaxies that belong to groups and those that
are relatively isolated. We also note the significant evidence of
galaxy-galaxy merger among the high and intermdiate mass
galaxies in poor groups, as evident from their luminosity
functions  
\citep{miles04,miles06}.

We divided our galaxy sample into
those that are a member of a group
(defined to be a member of a 2PIGG group \citep{eke2004} that
has 4 or more members), and 
those that are not
(isolated galaxies or those that
are members of 2PIGG groups of fewer than 4 members).
The majority of the galaxies (4921)
in our sample are part of groups
(or clusters) according to this definition.

The mean
value of $\eta$
parameter, and the fraction of passive galaxies, as a function
of their distance from the nearest cluster are plotted in
Fig.~\ref{sfrg}, 
which shows that the peak in mean $\eta$ 
occurs in groups and non-group galaxies at approximately the same
cluster-centric distance, 
but the non-group sample generally has a higher
fraction of star-forming galaxies.

\begin{figure}
\centerline{
\psfig{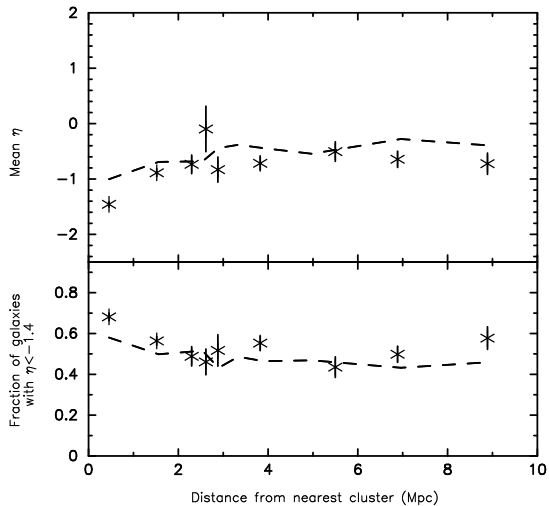} } 
\caption[Mean $\eta$ of filament and
  field group galaxies as a function of distance from the nearest
  cluster.]  {\footnotesize{ (a, Top) 
  The mean value of the parameter
  $\eta$, as a function of distance from the nearest
  cluster, for galaxies belonging to 
  2PIGG groups (with $\ge 4$ members)
   in the 2dFGRS. 
  Of these,  group galaxies that are also members of 
  the 52 filaments described in this work,
  are shown as crosses, while galaxies
  that are part of groups elsewhere in 
  the entire 2dFGRS shown as a dashed
  line.
  (b, Bottom) The fraction of galaxies (same symbols) with
  $\eta<-1.4$, as a function of distance from nearest cluster.}
\label{sfrg2df}}
\end{figure}

Fig.~\ref{sfrg2df} 
compares these same quantities for galaxies that
belong to groups that are part of filaments to similar group galaxies
elsewhere in the 2dFGRS.  The dashed line represents the variation of
mean $\eta$, and passive galaxy fraction as a function of distance
from the nearest cluster, of all galaxies belonging to groups (2PIGG,
$N\!\ge\! 4$) in the 2dFGRS.  The plot suggests that the star
formation rate is more or less uniform in all group galaxies
irrespective of their distance from the nearest cluster, except for
those galaxies within the virial radii of clusters. The crosses with
error bars represent the subset of these galaxies that belong to
groups that are part of the 52 ``clean sample'' supercluster filaments
studied in this work.  There is some indication that the trend we find
of a peak in SFR between 2--3$\, h_{70}^{-1}$~Mpc is seen in the group
galaxies that belong to filaments, indicating that the presence of the
filaments influences star formation properties in galaxies that belong
to groups in this region.

\section{Discussion}
\label{sec:dis}

At the outset, we mentioned the recent 
discovery of several instances of rapidly starbursting galaxies 
near the virial radius of rich clusters, with 
disturbed morphology
suggesting strong interactions
\citep[\eg][]{sun02,moran2005,reverte07}.
Does this occur at the periphery of every
rich cluster, as galaxies are accreted from their surroundings as part
of the hierarchical build-up of clusters? Does the incidence of such
objects depend on the nature of the environments of clusters, and on
the properties of the galaxies themselves? How common and important is
this effect in the life of a typical galaxy, formed on the filamentary
network of baryons as they are assimilated in the larger-scale scheme
of the Universe?

Here we undertook a statistical study of this effect
from a sample of clusters that can be seen to be fed by 
supercluster-scale filaments, from a substantial redshift survey
in the nearby Universe. In order not to be confused by superposition effects,
we selected a sample of ``clean'' 
inter-cluster filaments which do not have other
structures superposed on them, and studied the star formation
properties of galaxies as a function of position on
these filaments and in clusters at their extremities.

We collectively stack galaxies in all of the chosen filaments, and
look at the mean star formation rate, as well as the fraction of
passive galaxies, as a function of the 
the distance of each galaxy from its nearest
rich cluster (see Fig.\ref{sfr1}).
As expected, the SFR declines from the far outreach of the filament
filament environment to the
galaxies in the cluster core. This reveals the effect of a
physical mechanism
that quenches star formation progressively as a galaxy
approaches the core of the cluster potential well
\citep[e.g.,][]{balogh98,kauf2004,pimbblet2006,nolan07}.
However, as in our earlier study in the Pisces-Cetus supercluster
\citep{porter2007}, we find the evidence for a sharp burst of star
formation between $\sim$2--3~$h_{70}^{-1}$ Mpc, which is about 1.5--2
times the typical virial radius, from the centre of the nearest rich
cluster.

A typical disk galaxy formed in the relative sparse environment of
a supercluster filament is expected to have 
reservoir of warm gas to
provide for continued star formation in the galaxy.
This gas can be in the range of temperatures 
$10^5\! - \! 10^6$~K, which would account for X-ray emission,
or it could be cooler and be undetected in the X-rays.
Some evidence of such warm haloes has been
found around nearby disk galaxies from Chandra observations
\citep[e.g.,][]{pedersen2006}.

In their analysis of redshift catalogues, \citet{einasto91} found
  that the slope of the galaxy-galaxy correlation function is
  characteristic of near-spherical structures like groups and clusters
  for separations smaller than $3\,h_{100}^{-1}$ Mpc, beyond which it
  becomes more representative of filamentary structures. It is
  interesting to note that the characteristic cluster-centric distance
  at which the peak of star formation is seen in this study is roughly
  half of this spatial scale. Such a transition is apparent from
  Fig.~\ref{fill12}, where the Abell radius ($2.1\,h_{70}^{-1}$Mpc) is
  indicated, showing that on much larger scales, the predominant
  structures appear more filamentary than spherical.  This may
  indicate that such scales ($2\!-\! 3\,h_{70}^{-1}$Mpc) represent the
  zone of transition between the field from which galaxies are
  accreted into clusters, and the region in which the properties of
  the clusters become important in the evolution of galaxies.

  As a galaxy comes closer than this to the centre of the individual
  cluster, into which it is falling, the dark matter halo of the
  cluster begins to exert important tidal and dynamical influence.  A
  galaxy would possibly begin to encounter the hot intra-cluster
  Medium (ICM) of the rich cluster at this distance as well. As the
  gaseous halo of the galaxy interacts with the hot ICM, the galaxy
  may lose its reservoir of gas through evaporation and possibly ram
  pressure stripping.  This will lead to a steep decline in the SFR of
  galaxies within $\sim 1\,h_{70}^{-1}$~Mpc of the centre of the
  cluster, a trend that is seen in all relevant studies, including
  ours.  Where this process of ``strangulation'' of star formation
  becomes important, galaxy-galaxy harassment \citep{moore96} will
  also be an important effect. However, since most of the gas
  reservoir of the closely-interacting galaxies has been removed by
  this stage, star formation will not be induced by the harassment,
  and the remaining gas of the galaxy will continue to be stripped,
  leading to the observed steep dip in SFR in the first two bins in
  Fig.~\ref{sfr1} from the centre of each cluster.

However, approaching the outer regions of the cluster along a
filament, the infalling galaxies would experience close interactions
among themselves, even before they experience any significant
influence of the gravitational potential of the cluster, and that of
its ICM, on their rate of star formation.  Particularly for galaxies
falling along crowded filaments into a cluster, the local density of
galaxies would already have begun to increase rapidly, and it could be
appreciable at distances as large as twice the virial radius.
Therefore, the most likely cause for enhanced star formation that we
observe in a narrow range of cluster-centric distances between
$\sim$2-- 3$\, h_{70}^{-1}$~Mpc, would most likely be due to
galaxy-galaxy harassment, which is a rapidly acting process, working
efficiently in crowded environments. This effect would occur in
addition to the general trend of decrease in SFR towards the core of
the cluster, as found in galaxies elsewhere.  These galaxies, still at
relatively large distances from the cluster centre, yet to encounter
the hot and dense ICM of the cluster, would have not had their gas
stripped or evaporated, yet the close interaction with other galaxies
flowing into the same cluster would lead to density fluctuations in
their gaseous ISM, resulting in bursts of star formation.

In numerical
simulations, galaxy-galaxy harassment is seen to be a rapid effect
\citep[e.g.,][]{moore1999}. This accounts for the narrow range over
which the peak in the SFR is seen, since the effect depends on both
the existence of substantial fuel for star formation (at a substantial
distance from the cluster core), as well as sufficient external
influence (due to strong interactions from nearby galaxies) which
would act as trigger.  Furthermore, the star forming galaxies will
fall into the cluster with already elevated gas temperatures, which,
when they encounter the ICM, will make them more prone to evaporation
and stripping, which would steepen the gradient of the rapidly
decreasing SFR.  This effect may have been directly seen in the case
of a starburst galaxy in a group \citep{rasmussen2006}.

Within the virial radius of the cluster, the ``backsplash'' effect
\citep{gill2005,rines05},  resulting from a 
galaxy being on an oscillating orbit 
at the bottom of the potential well at the core of a cluster, is
expected to be strongest at about $\sim$1~Mpc from the centres of
clusters, and  thus is not likely to have a strong effect on the peak
in SFR seen in our sharp peak in star formation, which occurs at
larger cluster-centric distance.

\citet{moran2005} 
observe a similar sharp peak in SFR (evident from redshifted
[OII] emission) in the outskirts of the rich cluster CL0024 at $z\!=\!
0.4$, at $1.8\, h_{70}^{-1}$~Mpc from the cluster centre, where galaxy
harassment is cited as a possible cause. All of their starburst
galaxies are dwarfs ($-20\!<\! M_B$)
At lower redshift, a photometric study \citep{haines06a}
of the nearby Shapley supercluster
\citep{shap89,shap91} reveals an excess of 
 blue star-forming dwarf galaxies ($M_R\!>\! -18$)
at $\sim 1.5\, h_{70}^{-1}$~Mpc from the centre
of the rich clusters in the core of the supercluster.
While it is encouraging to
observe similar effects in other studies using different observables,
the SFR peak found in the case of the Shapley supercluster or  
CL0024 is a factor of $\sim$1.5 closer to the cluster
core than the distance at which the SFR peak is found in our study.
It is possible that the galaxies in 2dFGRS filaments
experience the effect of galaxy-galaxy harassment at a larger distance
from the cluster centre, since the accretion of galaxies in this case
occurs along relatively narrow filaments, resulting in similar galaxy
densities further out than would be seen in the case of clusters where
galaxies are being accreted from all directions.

Fig.~\ref{sfr1dg} shows 
that the  galaxies 
of lower luminosity ($-20\!<\! M_B \!\le\! -17.5$)
have a higher SFR than the giant galaxies
at all distances from the cluster core.  This is consistent with the
results of other studies \citep[e.g.,][]{haines06b,moran2005}.  This again
provides more evidence for galaxy-galaxy harassment being the dominant
process in producing  
the rapid burst in star formation, since dwarf galaxies are expected
to be more vulnerable to galaxy harassment than
giants.  Furthermore, the
small merger cross section of dwarf galaxies would make the merging of
galaxies an unlikely cause of the dwarf dominated peak in SFR.

Finally, we investigated whether a galaxy belongs to a group
(2PIGG group, $N\ge 4$) 
or not makes a difference to their star formation properties
 as they fall into clusters.  Within the filaments, the SFR peak
 occurs in the same place for group galaxies as in non-group galaxies, 
 though the peak is less pronounced in the former category.
We find that in the filaments,  the mean $\eta$ of the non-group
galaxies is  higher than that of the group galaxies.  The justification of
this could be that 
galaxies in the groups will have been interacting with each other
within the group environment long before they reach the infall regions
of the clusters. They will therefore have had some of their gas
already stripped off by galaxy-galaxy and galaxy-IGM 
interactions, by the time they reach their current position on the
filament. Therefore, the group galaxies will not have as much of a
fuel reservoir for continued star formation as the non group galaxies
and so will have lower mean SFR.

At this stage, we  admit to 
the caveat in the use of the $\eta$ parameter as a proxy for current
SFR, since  the principal component analysis used in obtaining
the $\eta$ parameter will not have removed contributions from active
galactic nuclei (AGN). It therefore remains a possibility that part of
enhancement in mean $\eta$ is due to AGN activity rather than
conventional star formation, consistent with the results of
\citet{ruderman2005}, who find a prominent spike in the number of
X-ray AGN in clusters at $\sim 2.5 \, h_{70}^{-1}$~Mpc. However, not
all X-ray detected AGN have the usual emission lines in their optical
spectra \citep{shen2007}. Clearly, further studies should involve
direct estimate of star formation rates from well-calibrated spectral
line features.


\section{Conclusions}
\label{sec:con}

From a  study of galaxies
belonging to three inter-cluster filaments in the
Pisces-Cetus Supercluster, 
\citet{porter2007} had found 
that, in addition to the (expected) systematic decline in the star
formation rate from the filament environment on the periphery of a
cluster to the cluster core, there is a peak, representing a burst of
star formation, over a narrow range of cluster-centric distance, at
$\sim$2--3 $\, h_{70}^{-1}$~Mpc (about 1.5--2 times the cluster virial
radius) from the centres of the clusters at the ends of each
filament. However, these results were based on a few hundred galaxies
in only three filaments, in one Supercluster.

In this work, we have repeated a similar analysis with a much larger
sample of galaxies (N=6,222), in 52 different inter-cluster filaments,
chosen from a redshift survey of a substantial fraction of
the $z\!<\!0.2$ Universe (from the 2dFGRS).

We chose a sample of ``clean'' filaments from a larger catalogue 
of supercluster filaments, requiring the filaments to connect two
rich clusters of galaxies and be at least 6$h_{70}^{-1}$~Mpc
long with no intervening clusters within this distance. The filaments
were chosen solely from their morphology without any
knowledge of the star formation properties of the galaxies belonging to them.

For galaxies belonging to the ``clean sample'' filaments, 
we have also looked at similar dependence on
distance of the fraction of passive galaxies, in each distance bin.
In doing so, we have also noted whether these galaxies are giant ($M_B
\!\le\! -20$) or a dwarf, or are a member of a group or a cluster.
Where possible, we have compared these results to similar properties
of galaxies elsewhere in the 2dFGRS galaxy redshift survey.

We confirm that in the cores of rich clusters, the star formation rate
in a galaxy is very low, irrespective of its mass, by showing that the
value of mean $\eta$ and the fraction of passive galaxies falls to its
lowest value for both giant and dwarf galaxies in the cores of
clusters. As one moves away from the cores of clusters along the
filaments, there is an increased activity of star formation, peaking
at approximately 2--3$ \, h_{70}^{-1}$~Mpc, at approximately 1.5 times
the virial radius of the clusters. This peak in star formation in
filament galaxies is seen to be mostly due to dwarf galaxies
($-20\!<\! M_B \!\le\! -17.5$).

The sudden enhancement
in the star formation rate witnessed 
in our sample (see Fig.\ref{sfr1})
is a rapidly acting process such as galaxies experiencing the
intracluster medium for the first time or galaxy harassment.  As
galaxies just start to encounter the ICM they will not yet have had
their gas stripped or evaporated, but the interaction with the ICM
will lead to tidal shocks leading to a burst of star
formation. Similarly, with their gas still intact, galaxy harassment
may lead to density fluctuations in the gas which may trigger a burst
of star formation. In the infall region of the cluster at a few $\,
h_{70}^{-1}$~Mpc, the galaxies will just have started to accelerate to
the velocities needed for galaxy harassment to become an important
factor. The sharpness of the peak in SFR is another indicator that
galaxy harassment may be important, since \citet{moore1999} show that
harassment is a rapid effect.

Within the filaments, the abrupt enhancement in star formation occurs
in the same place outside the virial radius of the cluster, for
galaxies belonging to 2PIGG groups, as in non-group galaxies, though
the peak is less pronounced in galaxies belonging to groups.  We
suggest that this due to the close interaction of galaxies within
groups causing some of the gas content of the galaxies to be stripped
away, before they experience the galaxy-galaxy harassment that is
responsible for the enhancement, even before they fall into the
cluster.

Using the $\eta$ parameter, which is derived from optical spectra,
and known to correlate with the current rate of star formation of a
galaxy, we have been able to show that on filaments feeding clusters, the
current star formation rate of a galaxy, particularly that of a dwarf,
is likely to undergo a sudden burst due to the close interaction with
other galaxies falling into the cluster along the same filament, if
the interaction occurs before the gas reservoir of the galaxy gets
stripped off due to other effects. The direct detection of these
galaxies, often excluded from narrow-field studies of the cores of
clusters, should be possible from the detection of dust from {\it
Spitzer} observations. From deep X-ray observations, it would also be
interesting to study how the group environment around each galaxy
(characterised by the hot IGM) affects its star formation rate, as the
group is assembled into the rich cluster.

Finally, we would also like to point out
that others have found similar starburst and IR-luminous
galaxies in the inner parts of certain
rich clusters as well 
\citep[e.g.][]{keel03,duc2004,cortese07}. 
The use of directly measured star formation parameters, instead of
proxies like the $\eta$ parameter will also reveal 
\citep[e.g.][]{mr08} 
interesting aspects of galaxy properties in such environments.
As larger samples of
such cases emerge, it would be interesting to investigate
whether these galaxies are actually near the virial radius
of the cluster, but seen
in projection to be closer to the centre, or if there are other 
physical effects, applying to a category of galaxies,
that protect the fuel for star formation
from being stripped as the galaxy falls into the core of a cluster.
Numerical simulations with appropriate resolution
and detailed 
hydrodynamical modeling of the
gas--galaxy interaction
\citep[e.g.][]{berrier08}, 
would be necessary to further investigate the relative importance 
of various physical processes affecting galaxy evolution over this
range of distances.


\section*{Acknowledgements}
SCP would like to thank  PPARC, UK for support through a 
research studentship.
SR would like to thank Christine Jones, Bill Forman and Trevor Ponman
for helpful discussions and encouraging noises. 
KAP acknowledges partial support through a UQ ResTeach Fellowship.
We would also like to thank the referee for helpful comments.


\label{lastpage}
\end{document}